\documentclass[a4paper,11pt]{article}
\pdfoutput=1 

\usepackage{jinstpub} 

\title{\boldmath Operational Experience with the ALICE Pixel detector}

\author[a]{A. Mastroserio,\note{Corresponding author.}}

\affiliation[a]{Physics Departiment of the Bari University and INFN Bari,\\V. Orabona 4, Bari, Italy}

\emailAdd{Annalisa.Mastroserio@ba.infn.it}

\abstract{The Silicon Pixel Detector (SPD) constitutes the two innermost layers of the Inner Tracking System of the ALICE experiment and it is the closest detector to the interaction point. As a vertex detector, it has the unique feature of generating a trigger signal that contributes to the L0 trigger of the ALICE experiment. The SPD started collecting data since the very first pp collisions at LHC in 2009 and since then it has taken part in all pp, Pb-Pb and p-Pb data taking campaigns. This contribution will present the main features of the SPD, the detector performance and the operational experience, including calibration and optimization activities from Run 1 to Run 2.}


\arxivnumber{999.999} 

\collaboration[c]{on behalf of the ALICE collaboration}

\proceeding{Pixel 2016, 8$^{\text{th}}$ on Semiconductor Pixel Detector for Particles and Imaging\\
  5-9 September\\
Sestri Levante, Italy
}

\begin{document}
\maketitle
\flushbottom

\section{Introduction}
\label{sec:intro}
The ALICE experiment at CERN is devoted to the study of a new state of matter that is formed in ultrarelativistic heavy ion collisions, namely the Quark Gluon Plasma (QGP). Its Inner Tracking System (ITS) is the closest detector to the beam pipe and it has been designed to cope with 50 particles per cm$^{2}$ in central Pb-Pb collisions. The ITS has a barrel geometry and it is composed by six cylindrical layers of silicon detectors, whose technologies and layout were optimized to improve tracking efficiency and impact parameter resolution down to very low momenta. The two innermost layers are constituted by Silicon Pixel Detectors (SPD), the middle layers are constituted by Silicon Drift Detectors (SDD) and the two outermost layers are constituted by double-sided Silicon Strip Detectors (SSD).  The SPD is the detector closest to the interaction region and therefore plays a crucial role for primary vertex reconstruction and secondary vertex resolution as well as for the event triggering. Indeed it has the unique feature of being a vertex detector that was designed also to provide a L0 trigger to the ALICE experiment.

\section{Detector Overview}
\label{sec:detector}
\begin{figure}[htbp]
\centering 
\vspace{-6 cm}
\includegraphics[width=\textwidth]{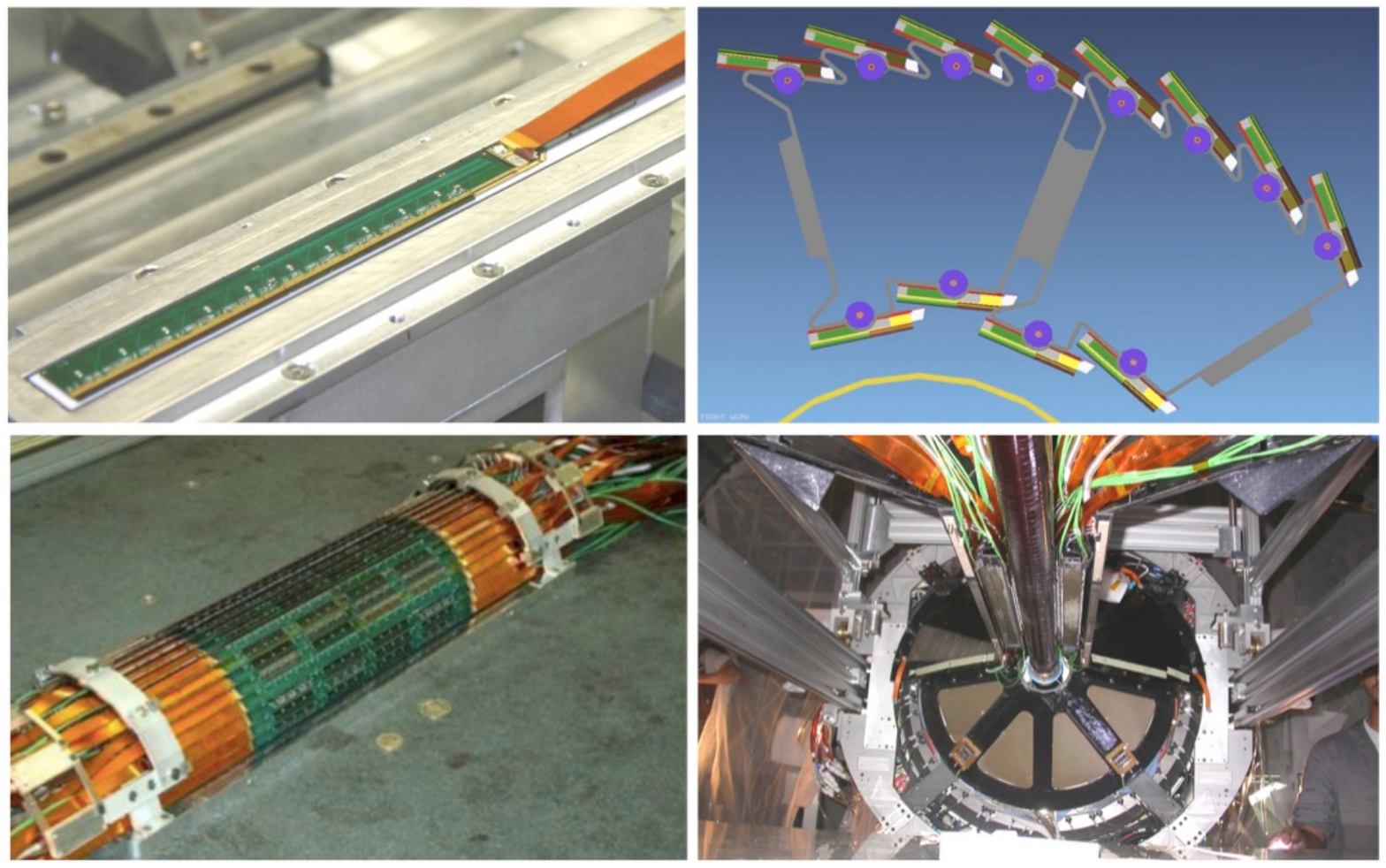}
\vspace{-7 cm}
\caption{\label{fig:overview}  Upper row shows the single half-stave and a schematic view of a sector. Bottom row shows the integration of 60 half-staves in a half barrel and its installation around the beam pipe.}
\end{figure}

The Silicon Pixel Detector ~\cite{spd} is the innermost ALICE detector positioned at 5 mm from the beam pipe. It has a barrel geometry and its innermost and outermost radii are 3.9 cm and 7.8 cm respectively that translates in a pseudorapidity coverage of  $\eta$<2 and $\eta$<1.4 respectively. In the transverse plane it is segmented in 10 sectors which cover 36$^{\circ}$ each in the azimuthal angle $\phi$. Each of them contain 12 half-staves, the smallest functional pieces of the detector, for a total of 1200 readout chips and 9.8 x 10$^6$ pixels overall.  Figure ~\ref{fig:overview} shows from the upper left panel to the bottom right panel the single half-stave, a schematic view of two sectors, the  integration of 60 half-staves in an half-barrel and the installation of the top half-barrel around the beam pipe. 
\vspace{0.5 cm}

The total power consumption of the SPD front-end electronics is about 1.3 kW, with a power consumption of 0.5 W per cm$^2$. The reduced material budget and the high density of the components is such that the detector temperature would increase at a rate of 1$^{\circ}$C/sec in case of a cooling failure. The required cooling efficiency is reached by an evaporative C$_4$F$_{10}$-based system with cooling capillaries running underneath the front-end chip, that allow to keep the operating temperature at around 30$^{\circ}$C.  The pixel chip can be configured by 42 parameters, remotely adjustable through 8-bit DACs such as current and voltage bias references, trigger delay, global threshold voltage and leakage current compensation.
Furthermore each chip can generate a digital binary signal, namely the FastOr signal provided by a dedicated circuitry.  Such signal is true if at least one pixel is active within the chip matrix. The 1200 FastOr signals are sent to the PIxel Trigger (PIT) ~\cite{pit} that was designed to process a predefined set of 10 algorithms to contribute to the L0 trigger decisions. A challenging requirement was to reach a latency of 800 ns  : given the various transmission times on the optical fibers, the processing time allowed in the pixel trigger board is limited only to 25 nsec. The list of the trigger algorithms used by the PIT is shown in table~\ref{tab:algo}. The detector was commissioned in 2008 and allowed to record very first data with cosmics, the first collisions in 2009 ~\cite{pit} and collected data during all the LHC Run 1 period (end in March 2013) from pp to p-Pb and up to Pb-Pb collisions. Several interventions were done during the Long Shut Down 1 and the SPD was recommissioned with beams in May 2015 for the LHC Run 2 period. The restart was very smoothly and the performance similar to Run 1; in particular, the operation during the dedicated data taking with Pb-Pb at $\sqrt{s_{NN}}$ = 5.02 was very satisfactory, including the new trigger functionalities implemented during LS1 (first Long Shutdown). 

\vspace{0.5 cm}
\begin{table}[htbp]
\centering
\caption{\label{tab:algo} Trigger algorithms of the PIxel Trigger system (PIT). The par$_{TOT}$, par$_{INN}$, and par$_{OUT}$ are the threshold values for the total number of fired FastOr, the fired FastOr in the inner layer and the fired FastOr in the outer layer respectively. All the values are stored in a database in the experimental area and exported offline at each run.}
\smallskip
\begin{tabular}{|l|c|l|}
\hline
Output & Name & Algorithm\\
\hline
1 & Minimum Bias & (I+O)$\geq$ par$_{TOT}$ and I$\geq$par$_{INN}$ and O$\geq$par$_{OUT}$\\
\hline
2 & High Multiplicity 1& I$\geq$ par$_{INN}$and O$\geq$par$_{OUT}$ \\
\hline
3 & High Multiplicity 2& I$\geq$ par$_{INN}$and O$\geq$par$_{OUT}$ \\
\hline
4 & High Multiplicity 3& I$\geq$ par$_{INN}$and O$\geq$par$_{OUT}$ \\
\hline
5 & High Multiplicity 4& I$\geq$ par$_{INN}$and O$\geq$par$_{OUT}$\\
\hline
6 & Generalized topological trigger & based on tracklets \\
\hline
7 & Less Than & I$\geq$par$_{INN}$and O$\geq$par$_{INN}$ \\
\hline
8 & Spare background & O$\geq$I+ offset$_{OUT}$\\
\hline
9 & Background  & (I+O)$\geq$par$_{TOT}$ \\
\hline
10 & Cosmics &  Selectable coincidence \\
\hline
\end{tabular}
\end{table}
\vspace{1 cm}

\section{Configuration and operation}
\label{sec:Run1Run2}
\begin{figure}[htbp]
\centering 
\includegraphics[width=0.75\textwidth, angle=-90]{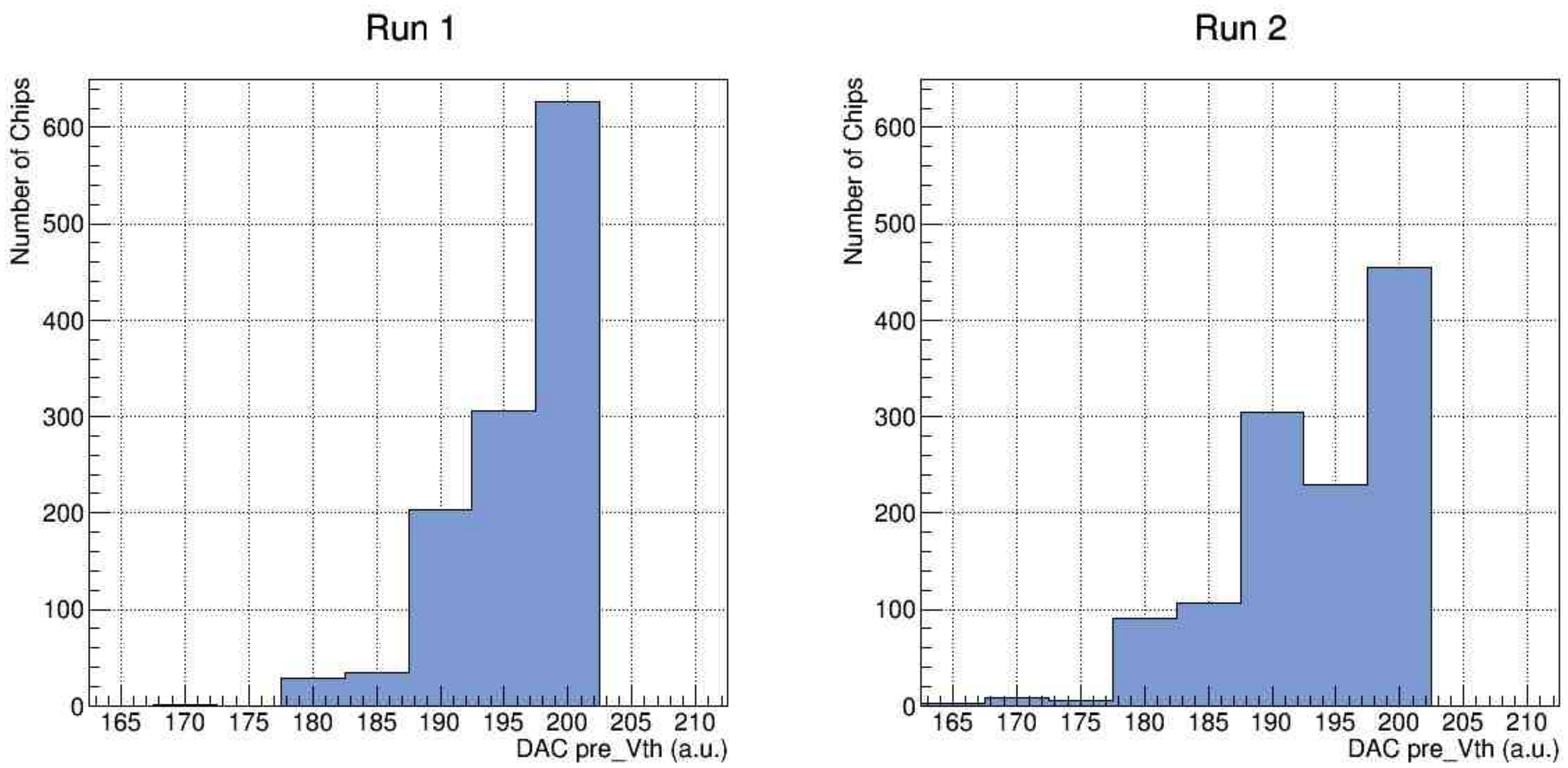}
\vspace{-1 cm}
\caption{\label{fig:preVth} Distribution of the preVth DAC of the 1200 chips in Run 1 (left) and Run 2 (right). The mean value of the two distributions lowered from 196 to 193. This result means a general increase of the thresholds (see text for more details).}
\end{figure}

\begin{figure}[htbp]
\centering 
\vspace{-1 cm}
\includegraphics[width=.7\textwidth, angle=270]{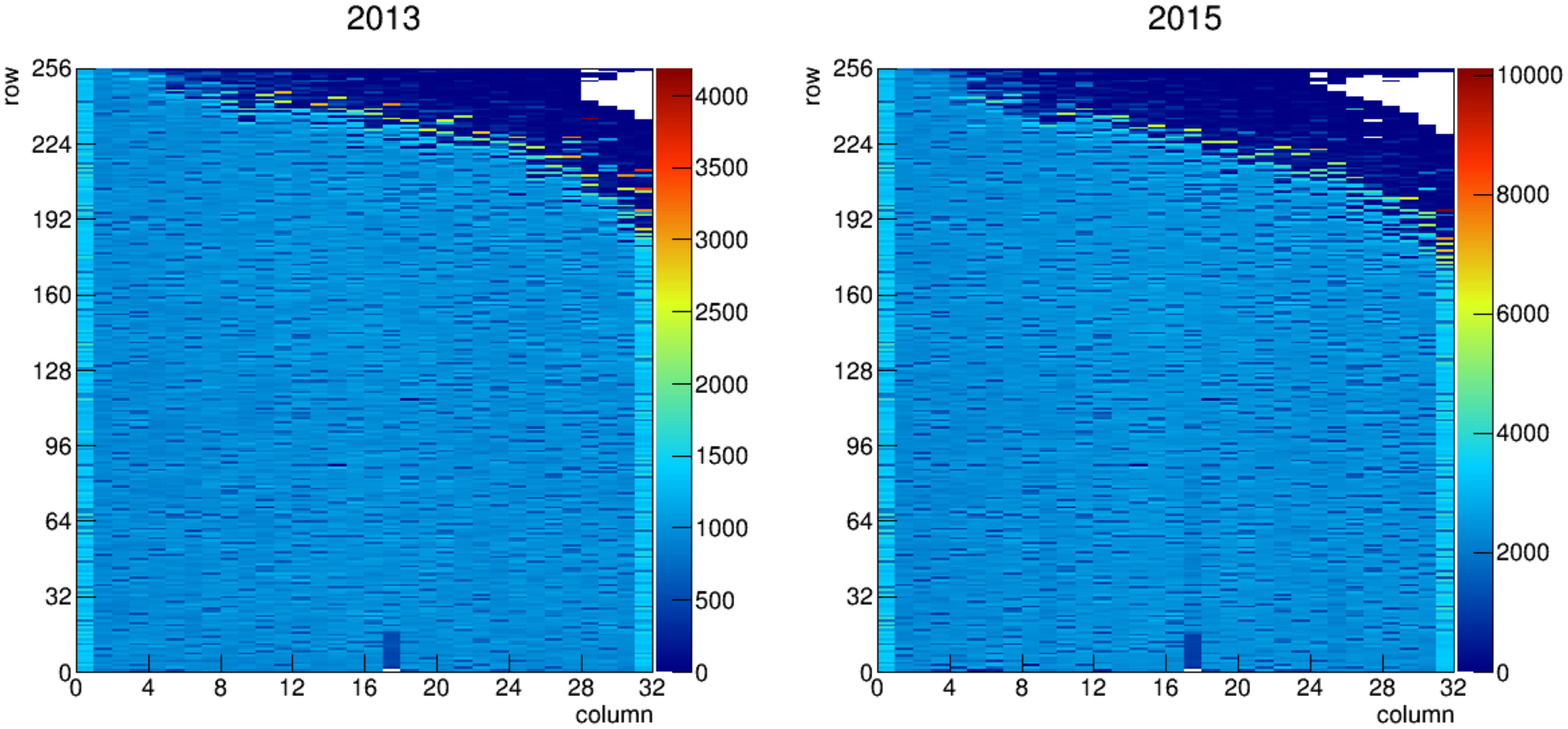}
\vspace{-1 cm}
\caption{\label{fig:chip} Pixel hits in one chip from Run 1  (p-Pb collisions) on the left and Run 2 (Pb-Pb collisions). The white area that corresponds to pixels with missing bump-bonding has slightly increases, the blue region is the transition region between such a dead area and the one containing pixels with working bondings. The two bands on the sides of the chip correspond to pixels having a wider size in z (625 $\mu$m instead of 425 $\mu$m).}
\end{figure}

The detector configuration was finalized in Run 1 and since then there detector performances were stable. It is worth mentioning that the detector operation in Run 1  was affected by a problem with the filters of the cooling system. Nevertheless after a very challenging operation of drilling the filters which were located in a position not easily accessible, the number of half-staves in the data acquisition reached 110 out of 120. The actual number of operational half-staves in Run 2 is 112 thanks to a further improvement of the cooling system provided by the introduction of a gear pump. The configuration settings of the SPD at the beginning of Run2 were the same as the ones used in Run 1, nevertheless a threshold campaign was performed to include the detector in the triggering of specific pp events. Run 2, with its the higher luminosity, needed more refined triggers therefore the trigger performances of three algorithms for dedicated collision topologies were exploited. Despite the increased number of bunches in each beam, the readout strobe of the detector was still kept to 300 ns (3 clock cycles), whereas the FastOr strobe was kept to the minimum value of 100 ns.\\
The  new luminosity scenario implied that the noise of the SPD trigger needed to be kept under control and to the level requested by the ALICE trigger coordination group. The 8 bit DAC that controls the threshold at the chip level is  called \textit{preVth} and figure ~\ref{fig:preVth} shows its distribution among the chips before and after the tuning. Given that a preVth value equals to 190 corresponds to 3100 electrons and that a value equals to 200 corresponds to 2500 electrons, the reduction of its value corresponds to a threshold increase. The mean value of the distribution lowered to 193 (was 196 in Run 1), therefore the average value of thresholds slightly increase,  although still very well below the amount of charge produced by a minimum ionizing particle traversing the 200 $\mu$m thick sensor. A consequence of this change was observed in the chip at the transition regions between bumb-bonded pixel and missing bump-bondings. The thermal cycles of the half-staves made the readout chip detaching from the sensor affecting the bump-bondings. Figure \ref{fig:chip} shows the pixel hits from one chip (chip 1 from the half-stave 2 of the halfsector 1) in the very last runs of Run1 (p-Pb collisions) and in Pb-Pb collisions from 2015. From the missing bump bonding area, the white area, it is clearly visible a transition region in blue to the efficient part of the chip (light blue). The pixels corresponding to that region are monitored periodically by means of a dedicated procedure. Within such a procedure the pixels that can be identified as inefficient: their fraction has changed from 2\% in Run 1 to roughly 4\%  in Run 2.\\

\vspace{0.5 cm}
As far as the noisy pixels are concerned, such pixels can be grouped in two main categories: the definitively noisy pixels and the ones that appear noisy during data taking. The first type of pixels is  found in self triggering mode assuming a threshold of 2\% of the events, the second ones rarely appear (they are usually one or few units maximum per run) and usually they are selected if they are active in more than 5 \% of the events. \\
After the latest threshold campaign the detector operation was stable and allowed to take data in the high luminosity regime of high energy pp collisions ($\sqrt s$=13 TeV) and high energy Pb-Pb collisions ($\sqrt{s_{NN}}$=5.02 TeV). 
The detector configuration that consists in the set of 42 parameters for the front end electronics and the trigger thresholds of the 10 algorithms, is stored in the Alice Configuration Tool (ACT) and it is checked automatically at each run. The stability fo the detectors is shown by the fraction of physics runs including SPD that was 93\% (86\% at the end of Run1) and the fraction of runs stopped due to the SPD being in error lowered to 2\% (4\% at the end of Run1). The main issues faced during the new data taking period were the loss of configuration of one half-stave and data format errors (missing header/trailer). In the first case the run can proceed and the misconfigured element is masked offline for analysis. In the second case, unless the data are missing for hardware reasons (loss of optical link connection), the run is not stopped and the problematic chips are further masked offline.
\section{New operation features in Run 2}
\label{sec:trigger}
The main novelty in the detector operation in Run 2 is the exploitation of its trigger capabilities. Most of the Pixel Trigger Algorithms are based on selecting events where the total number of fired FastOr in the inner layer, outer layer or both layers are above a predefined threshold (see table~\ref{tab:algo} for further details). Each  threshold is chosen according to the physics case of interest and it is stored in the ACT. Nevertheless in Run 2 a new algorithm based on event topology has been introduced and a new firmware was  deployed to this aim). Starting already from the beginning of 2015, the background rejection algorithm was tested to reject beam-gas interactions. In beam-beam collisions, in fact, the fired FastOr in the inner and in the outer layers is pretty much the same, so in case of beam gas interaction such a difference is far from being zero and the algorithm number 8 of table ~\ref{tab:algo} is suitable for triggering on background events. It was observed that the number of such triggers followed the trend of the Beam Line Monitoring detectors (BLM) outputs and allowed a rejection of 40 \% of background events using as an offset for the outer layer 20 FastOr. \\ 
\subsection{High Multiplicity trigger}
Recent physics results increased the interest in looking at different colliding system at the same mulitplicity, so in Run 2 a trigger on high multiplicity pp events was introduced. As far as the PIT is concerned, four identical algorithms are dedicated to trigger on high multiplicity events (see table ~\ref{tab:algo}). One of such trigger algorithms was configured such that the  thresholds on the number of chips of the inner and the outer layer were set to par$_{INN}$ = 0 and par$_{OUT}$ = 70 respectively. 
\vspace{0.5 cm}
\begin{figure}[htbp]
\centering 
\includegraphics[width=.5\textwidth]{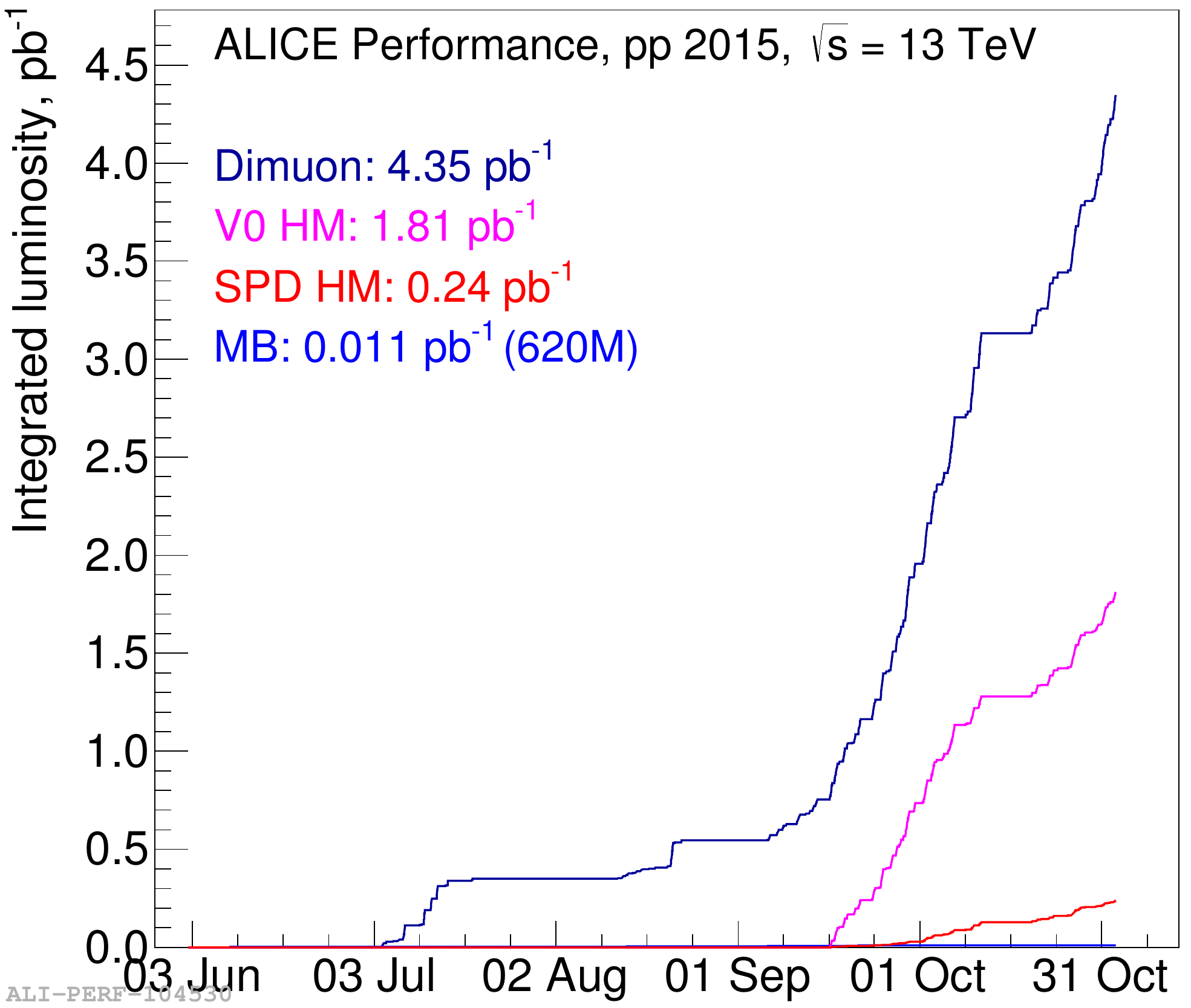}
\caption{\label{fig:highM} Integrated luminosity of pp events for Minimum Bias triggers and high multiplicity triggers (HM) as seen by the V0  and the SPD detectors. Their rapidity coverage is different, in particular the V0 acceptance is 2.8<$\eta|$<5.1 and -3.7<$\eta$<-1.7, whereas the SPD acceptance is limited to the central rapidity region  $|\eta|<1.5$ so an HM event is detected less frequently.}
\end{figure}
Figure ~\ref{fig:highM} shows the integrated luminosity collected by this trigger together with the HM trigger from the V0 detector \cite{v0}. Such detector has a different acceptance with respect to the SPD. In particular it triggers on particles produced at forward rapidities, whereas the pixel detector is able to trigger on particle produced at midrapidity. The data triggered by the silicon detector are under study by dedicated physics working group especially to disentangle true high multiplicity pp events from the contribution coming from the pile up of low multiplicity pp events  in the same recored event.

\subsection{Topological trigger}
\begin{figure}[htbp]
\centering 
\includegraphics[width=.65\textwidth, angle=270]{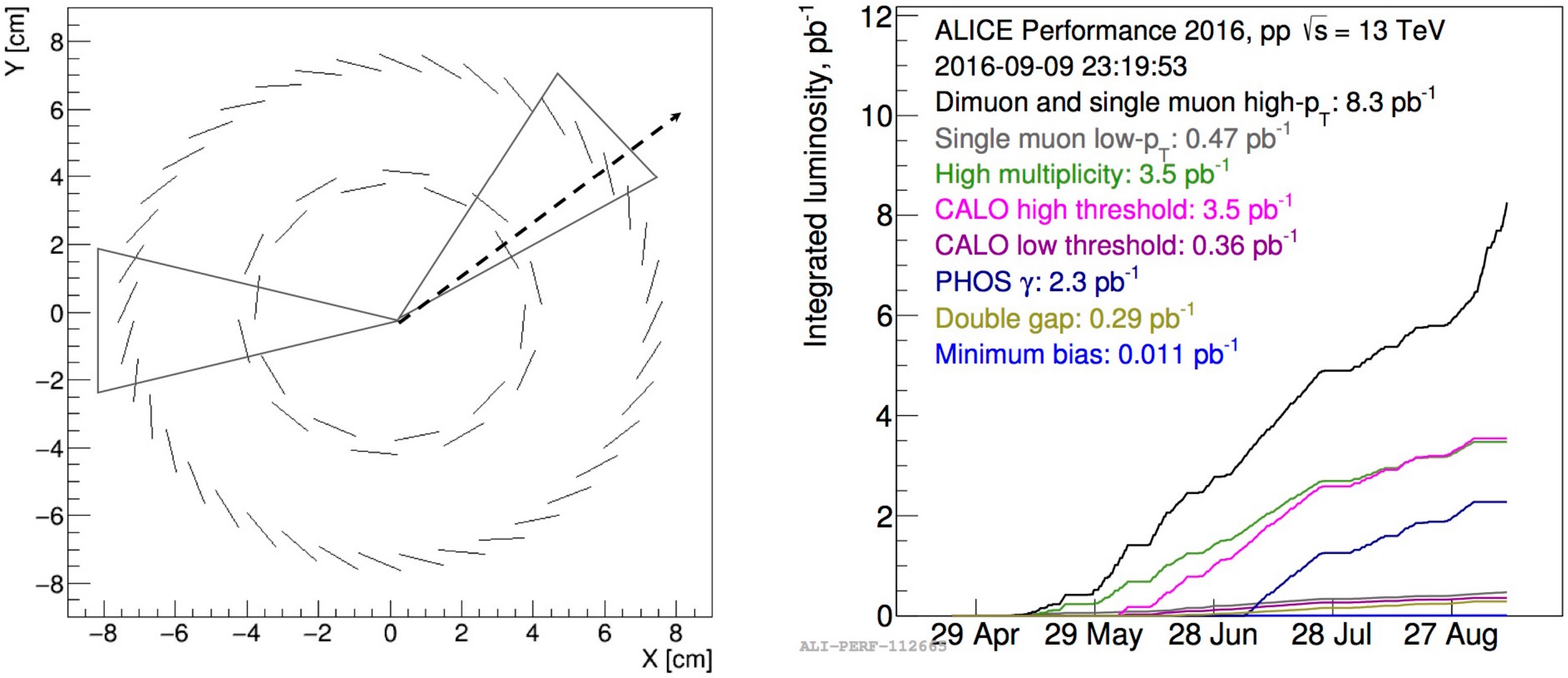}
\vspace{-2 cm}
\caption{\label{fig:dGap} Left panel : The dashed line represents a charged particle passing through the detector and firing two chips in two different layers. In general such chips are located at similar azimuthal angles. The cones, instead, are inside one sector and they represent the search area for tracklets of the topological algorithm. Right panel : Integrated luminosity of several ALICE trigger types used in Run 2 (MB, HM, etc.). The topological trigger one is shown  by the dark yellow line and named \textit{Double gap}.}
\end{figure}
\vspace{0.5 cm}
A new algorithm was introduced in Run 2 to trigger on double gap diffractive pp events. The basic concept that this algorithm is based on is the \textit{tracklet} concept: a cluster created by a charged particle in a chip of the inner layer and a cluster created by the same track in a chip of the outer layer have similar azimutal angles (see dashed line in the left panel of Figure ~\ref{fig:dGap}). A tracklet, then, is the association  of two chips in two different layers that are close in r$\phi$, that is translated in close FastOr belonging to two different layers in the same sector. The algorithm is able to identify a tracklet within each of the cones shown in the left panel of figure ~\ref{fig:dGap}). The algorithm subsequently can select also the opening angle between two cones containing tracklets. Such features allow to select specific event topologies. Right panel of Figure ~\ref{fig:dGap} shows the integrated luminosity of this trigger. The data taken from such triggered collisions are already under study.
\vspace{0.5 cm}

\section{Conclusions}
The SPD is recording data since the very beginning of Run 1 which saw the very first LHC collisions in 2009 and it steadily takes part in the physics data taking  in Run 2 started in June 2015. Its latest configuration is even better than the one in Run 1 : slight changes were applied in the thresholds to reduce the trigger noise according to the ALICE trigger group requests. Its operation didn't change from the previous data taking period and so far it revealed to be very stable in taking data from high luminosity pp collisions to Pb-Pb collisions at $\sqrt{s_{NN}}$=5.02 TeV. The main difference in Run 2 consisted in the exploitation of the SPD triggering capabilities with the High Multiplicity trigger as well as the very recent topological trigger for the dedicated pp event topologies. The data collected by both triggers are under study by dedicated physics working groups.


\end{document}